# Universal scaling of electrostatic effects of a curved counter-electrode on the emitter field enhancement

Thiago A. de Assis[1, 2, *] and Fernando F. Dall'Agnol[3, †]

[1]*Instituto de Física, Universidade Federal da Bahia, Campus Universitário da Federação,*
*Rua Barão de Jeremoabo s/n, 40170-115, Salvador, BA, Brazil*
[2]*Instituto de Física, Universidade Federal Fluminense,*
*Avenida Litorânea s/n, 24210-340, Niterói, RJ, Brazil*
[3]*Department of Exact Sciences and Education, Universidade Federal de Santa Catarina,*
*Campus Blumenau, Rua João Pessoa, 2514, Velha, Blumenau 89036-004, SC, Brazil*

Experiments on field electron emission from single-tip nanoemitters have typically been carried out using a counter-electrode with a finite curvature radius $R$, positioned at a distance $d_{\text{gap}}$ from the emitter's apex. The effects of the counter-electrode's curvature on the apex field enhancement factor ($\gamma_{\text{Ca}}$) of the emitter are still not understood. In this Letter, we theoretically explore how the apex field enhancement factor of an emitter, represented by a hemisphere on a cylindrical post (HCP) with apex radius $r_{\text{a}} = 50$nm, is influenced by the curvature of a spherical-shaped counter-electrode. Importantly, our results show that for HCPs with sharpness aspect ratios typically between $10^2$ and $10^3$, there is a universal scaling such that $\gamma_{\text{Ca}} = \gamma_{\text{Pa}} \Psi(R/d_{\text{gap}})$, where $\gamma_{\text{Pa}}$ represents the apex field enhancement factor for the emitter assuming a planar counter-electrode, and $\Psi(R/d_{\text{gap}})$ is a universal scaling function such that $\Psi \sim 1$ for $R/d_{\text{gap}} \gg 1$ and $\Psi \sim (R/d_{\text{gap}})^\alpha$, with $\alpha$ close to unity, for $R/d_{\text{gap}} \ll 1$. These findings help partially explain discrepancies observed in orhtodox field electron emission experiments, who reported that the effective $\gamma_{\text{Ca}}$ values extracted from the current-voltage characteristics of single-tip carbon nanotubes typically underestimate the theoretical $\gamma_{\text{Pa}}$ values when $R \sim d_{\text{gap}} \gg r_{\text{a}}$, a trend that is predicted by our results.

Field electron emission (FE) from pointed nanostructures, including carbon nanotubes (CNTs), has attracted significant attention due to its potential applications in nanotechnology and as an electron source [1–7]. Theoretical studies in this field often rely on simplified models [8–13], which neglect effects such as resistivity, temperature variations, and material-specific emission properties. While these assumptions abstract physical complexities, they enable precise solutions to Laplace's equation for electrostatic fields, offering a clear framework for understanding the relationship between emitter geometry and electrostatic field enhancement.

In a FE system, a nanostructure can be classically modeled as a hemisphere on a cylindrical post (HCP) [14, 15], with a total height $h$ and an apex radius of curvature $r_{\text{a}}$ (thus defining the sharpness aspect ratio $\sigma_{\text{a}} \equiv h/r_{\text{a}}$), standing on one of a pair of separated parallel planar plates with a lateral extent much larger than the plate separation $d_{\text{sep}}$. This arrangement is known as the parallel planar plate (PPP) geometry [15]. In this context, for a given inter-plate voltage, $\Phi_{\text{P}}$, the applied field $E_{\text{P}}$ is defined as $-\Phi_{\text{P}}/d_{\text{sep}}$. Therefore, the apex plate-field enhancement factor (FEF) is defined by $\gamma_{\text{Pa}} \equiv E_{\text{a}}/E_{\text{P}}$, where $E_{\text{a}}$ is the local electrostatic field at the apex of the post. The subscript "P" here is used to indicate that the counter-electrode is planar.

However, in the few representative experiments in the literature that describe the FE of a single-tip field emitter, the counter-electrode is curved [16–19]. In this case, the effects of the counter-electrode curvature radius $R$ on the emitter's apex FEF are unknown and will be unveiled in this Letter. As discussed below, our results provide a plausible explanation for the apparent discrepancies observed between related experimental findings and theory. Indeed, when extracting effective FEFs from the orthodox current-voltage characteristics [20] of a single emitter nanostructure in a FE system with a curved counter-electrode, the emitter's apex FEF, denoted in this case as $\gamma_{\text{Ca}}$ (the subscript "C" here is used to indicate that the counter-electrode in the FE system is curved) - which is equivalent to the previously defined $\gamma_{\text{Pa}}$ in the limit as $R$ tends to infinity - incorporates the effects of this curvature. In this case, for a given inter-plate voltage, $\Phi_{\text{C}}$, the applied field $E_{\text{C}}$ is defined as $-\Phi_{\text{C}}/d_{\text{sep}}$. Therefore, the apex plate-FEF when the counter-electrode is curved is defined by $\gamma_{\text{Ca}} \equiv E_{\text{a}}/E_{\text{C}}$. Importantly, our results show that there is a universal scaling relationship between $\gamma_{\text{Ca}}$ and $\gamma_{\text{Pa}}$, indicating that the ratio between these quantities depend solely on the ratio $R/d_{\text{gap}}$, where $d_{\text{gap}}$ is the distance between the spherical counter-electrode and the emitter's apex. This universality was observed for $10^2 \lesssim \sigma_{\text{a}} \lesssim 10^3$, a range of sharpness aspect ratios consistent with those reported in FE experiments.

At this point, it is useful to highlight that orthodox emission theory [21–23] relies on a well-defined framework of physical and mathematical assumptions, namely: (i) the theory is strictly one-dimensional and applies to a planar surface (although it remains reasonably accurate for surfaces with an apex radius $r_{\text{a}} > 30$ nm); otherwise, the standard Schottky–Nordheim barrier—which models correlation-and-exchange effects via a classical image potential energy and assumes the linearity of the applied electrostatic potential within the tunneling region—does not hold [24]; (ii) the electron energy dispersion relation, $E(k)$, scales as $E(k) \sim k^2$, where $k$ is its wave vector; and (iii) the wavefunctions that constitute the incident current are plane waves. However, in practice, FE scenarios rarely conform perfectly to this model, though certain specific conditions are believed to approximate it.

---

* thiagoaa@ufba.br
† fernando.dallagnol@ufsc.br

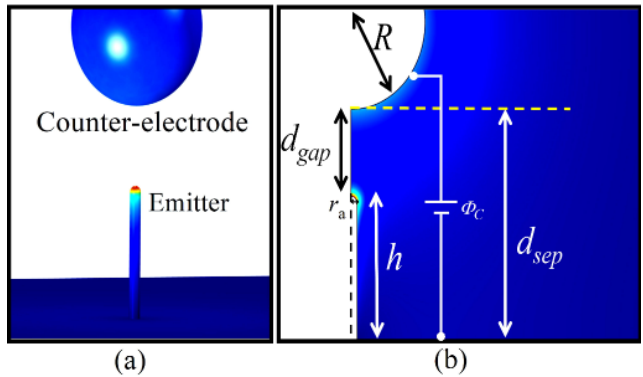

FIG. 1. (color online) (a) Three-dimensional visualization of a classical physical model containing an HCP emitter with $R/d_{\rm gap} = 1$, providing an overview of the counter-electrode close to the emitter's tip. The color map reflects the distribution of local electrostatic fields, with red (blue) representing stronger (weaker) fields. This system aims to represent the Scanning Electron Microscope (SEM) micrograph from [16] (see their Fig. 2) showing a CNT and a rounded counter-electrode with curvature radius $R \approx d_{\rm gap}$ (see the text for more details). (b) Axisymmetric two-dimensional representation of the FE system, which was used for our computational modeling. The relevant geometric parameters of this work and the boundary conditions are also shown.

Figure 1 (a) shows a three-dimensional visualization of a classical physical model containing an HCP emitter with $R/d_{\rm gap} = 1$, providing an overview of the counter-electrode close to the emitter's tip. This system aims to represent one of the experimental setups reported in Ref.[16], featuring a CNT and a rounded counter-electrode with curvature radius $R$, positioned at a distance $d_{\rm gap}$ from the emitter's apex (see Fig.1 (b) for the relevant geometric parameters of this work). Such setup is widely recognized FE experiment of individual CNTs grown by chemical vapor deposition in a scanning electron microscope. Interestingly, the measured current-voltage data yield nearly linear Fowler-Nordheim (FN) plots [25] that pass the orthodoxy test [20] (see Sec. 1 of the Supplementary Material). Therefore, for such nanoemitters, a constant effective $\gamma_{\rm Ca}$ [15] can be, in principle, extracted. This aspect is particularly notable, as the results are presented for CNTs with $r_{\rm a} = 7.5$ nm, a regime where orthodox field emission theory is not strictly applicable. At such small curvature radius, the second-order term in the Taylor expansion of the applied electrostatic potential $\Phi$ becomes significant, compromising its linearity with distance [24] (see Figs. S1, S2, and S3 in Sec. 2 of the Supplementary Material for our numerical calculations using the HCP model). Additionally, the specific energy dispersion relation of CNTs, which deviates from the $\sim k^2$ dependence, is expected to influence the shape of FN plots [26]. For single-walled carbon nanotubes (SWCNTs), field penetration within the caps further complicates the interpretation [27, 28] (see also the Supplementary Material of Ref. [29]). Nonetheless, recent quantum mechanical studies on SWCNTs indicate that, when properly defined within a polarization regime, the induced FEFs align well with those obtained classically using the HCP model, even for sub-nanometric apex radii [9, 10]. The experimental work of Bonard et al., which involves FEF measurements derived from current-voltage characteristics, passes the orthodoxy test—something initially unexpected. This highlights the need for further exploration of the relationship between FEFs extracted from FN plots for emitters with small apex radii and the actual characterization parameters of the emitters. While this issue lies beyond the scope of this Letter, we will explore the effects of the counter-electrode curvature on the apex FEF of a single-tip nanoemitter. Therefore, our simulation results focus on $r_{\rm a} = 50$ nm, where the theoretical framework remains robust.

As a relevant remark, it is worth noting that in the Bonard et al. experiment, the authors highlight two key findings: (i) simulations suggest that the shape of the counter-electrode does not significantly influence the obtained value of their $\gamma_{\rm Ca}$, as the apex radius of the CNT is much smaller than that of the counter-electrode; (ii) the experimental results for $\gamma_{\rm Ca}$ yield absolute values that overestimate the theoretical ones for most tubes, typically by a factor of 2. In this Letter, our results show that the experimental conditions used by the authors contradict point (i). As a result, part of the discrepancy found in point (ii) should be expected, as their experiment compares the effective experimental $\gamma_{Ca}$ with theoretical values obtained from simulations assuming a planar counter-electrode (i.e., $\gamma_{Pa}$). Next, we present the methodology used in this work, provide the rationale behind our claims, and discuss the corresponding results. In this study, we introduce two useful dimensionless variables, namely: $\Xi \equiv d_{gap}/r_a$ and $\Lambda \equiv R/r_a$.

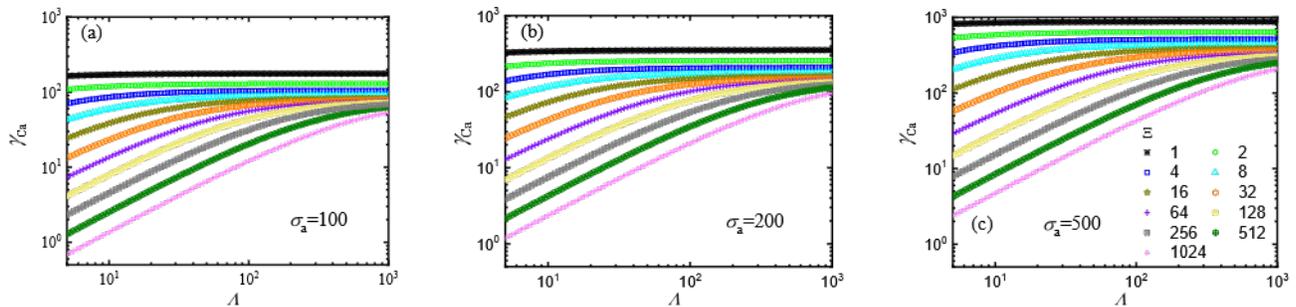

FIG. 2. (color online) $\gamma_{Ca}$ as a function of the variable $\Lambda \equiv R/r_a$ for $\Xi \equiv d_{gap}/r_a = 2^n$, with $0 \leqslant n \leqslant 10$, for (a) $\sigma_a = 100$, (b) $\sigma_a = 200$ and (c) $\sigma_a = 500$.

Figure 1 (b) illustrates the geometry and key features of the simulation domain, superimposed on the solution for the electrostatic field strength. The simulated FE system is modeled in a two-dimensional axisymmetric configuration, with the HCP emitter. A spherical counter-electrode is positioned above the emitter. To determine the $\gamma_{Ca}$, we need to solve the Laplace equation, $\triangle \Phi = 0$. This was numerically solved in a discretized domain $\Omega$ with $M$ nodes using the finite-element method (FEM) [15]. The electrostatic potential, $\Phi$, and the local electrostatic field distributions $[\mathbf{E}(\mathbf{r}') = -\nabla \Phi(r')]$ on emitter's surface were calculated. An approximate solution for the electrostatic potential was obtained in the form

$$\Phi^{approx} = \sum_{i=1}^{M} \Theta_i \Phi_i \approx \Phi, \qquad (1)$$

where $\Theta_i$ are basis functions (also called hat functions, with $\Theta_i = \delta_{ik}$, where $\delta_{ik}$ is the Kronecker delta) defined at the $i$-th node, and $\Phi_i$ are the nodal values of the electrostatic potential. The domain was composed of triangular elements, and each finite element was considered a first-order element with a triangular shape (defined by three nodes).

Next, FEM was applied to the Laplace equation with boundary conditions [shown in Fig. 1 (b)], yielding a matrix-vector equation $A\Phi_M = K$. The emitter plate is grounded, while the counter-electrode is held at a fixed voltage $\Phi_C$. The right-hand and top boundaries are treated as symmetry boundaries. We take advantage of the system's rotational symmetry to reduce computational complexity. The interior of both the emitter and the counter-electrode does not influence the $\gamma_{Ca}$ and is excluded from the domain. These boundaries are carefully placed to respect the minimum domain dimensions (MDD) [15, 30, 31] required to ensure that the error in the computed apex FEF does not exceed 0.1% compared to the case of a perfectly single-tip emitter.

The solution vector $\Phi_M$ (a column matrix with $M$ elements) contains the nodal values $\Phi_i$ of the approximate electrostatic potential solution. The matrix $A$ is a global $M \times M$ matrix with elements

$$A_{ij} = \int_\Omega (\nabla \omega_j \cdot \nabla \Theta_i) \, d\Omega, \qquad (2)$$

where $1 \leq (i,j) \leq M$. In the Galerkin method [32] used in this work, we have $\omega_j = \Theta_j$. The vector $K$ is a column matrix with $M$ elements, with

$$K_j = \int_\Omega \rho' \omega_j d\Omega. \qquad (3)$$

Since the charge density $\rho'$ is zero everywhere in $\Omega$, all elements of $K_j$ are null.

Next, let us discuss our theoretical results in more detail for a wide range of experimentally relevant parameters. Figures 2 (a), (b), and (c) show $\gamma_{Ca}$ as a function of $\Lambda$ for $\Xi = 2^n$, with $0 \leqslant n \leqslant 10$, for $\sigma_a = 100, 200$, and 500, respectively. The simulations were performed assuming $r_a = 50$ nm. For a given $\sigma_a$, the results show that when $\Xi$ is small, the variation of $\gamma_{Ca}$ with $\Lambda$ is not significant within the considered range of $\Lambda$. However, as $\Xi$ increases, an initial power-law regime is observed, i.e., $\gamma_{Ca} \sim \Lambda^\alpha$, when $\Lambda \ll \Xi$, before reaching the saturation regime, where $\gamma_{Ca}$ converges to a saturated FEF, $\gamma_S$. Interestingly, our numerical results show that $\gamma_S = \gamma_{Pa}$. In other words, $\gamma_{Ca}$ converges to $\gamma_{Pa}$ as $\Lambda$ becomes sufficiently large, for all values of $\Xi$. That is, in the saturation regime, the spherical counter-electrode interacts with the emitter as if it were nearly a flat plate. However, what does it mean $\Lambda$ to be sufficiently large to ensure that $\gamma_{Ca} \to \gamma_{Pa}$?

Before addressing the above question, it is instructive to compute the values of $\gamma_{Pa}$ and its variation with $\Xi$.

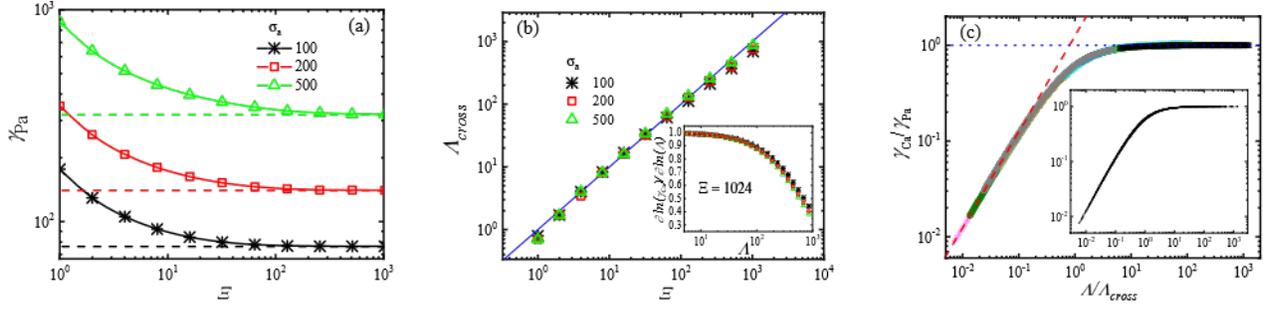

FIG. 3. (color online) (a) $\gamma_{Ca}$ as a function of $\Xi$ for $\sigma_a = 100, 200$, and $300$. The solid lines serve as a guide, connecting points that represent the numerical results. The dashed horizontal lines indicate $\gamma_{Pa}$ for each $\sigma_a$, as determined using the MDD extrapolation technique (see Table II of Ref. [31]), which assumes the counter-electrode is located at infinity. (b) $\Lambda_{\text{cross}}$ as a function of $\Xi$, for $\sigma_a = 100, 200$, and $300$ (see text for details). The solid line represents the first bisector. The inset shows, for the three values of $\sigma_a$ presented in Fig. 2, with $\Xi = 1024$ fixed, the local slope analysis. (c) Collapse of all curves (in log-log scale) shown in Fig. 2 (a), by replacing the variables $\gamma_{Ca} \to \Psi \equiv \gamma_{Ca}/\gamma_{Pa}$ and $\Lambda \to u \equiv \Lambda/\Lambda_{\text{cross}}$ (see text for details). The dashed and dotted lines represent a line with a slope equal to 1 and a constant function equal to 1, respectively. The inset shows the collapse of all curves presented in Figs. 2 (a), (b), and (c).

Our results are shown in Fig. 3 (a). For all the $\sigma_a$, a non-linear behavior is observed, characterized by a decrease in $\gamma_{Pa}$ with increasing $\Xi$, followed by a constant trend, which occurs in the limit where the counter-electrode is sufficiently far from the emitter. In this limit, we clearly observe that the values of $\gamma_{Pa}$ tend to those FEFs previously calculated with the MDD extrapolation technique, when the counter-electrode is assumed to be at infinity (see Table II of Ref. [31]). Next, referring to Fig. 2, we can observe that, for fixed values of $\Xi$ and $\Lambda$, $\gamma_{Ca}$ increases with the sharpness aspect ratio $\sigma_a$, which is also a consistent result. Despite this, the dependence of $\gamma_{Ca}$ on $\Lambda$ and $\Xi$ follows a similar trend for the studied values of $\sigma_a$, i.e., a power-law growth followed by saturation, as discussed earlier. This behavior suggests the existence of a characteristic $\Lambda$, denoted as $\Lambda_{\text{cross}}$, which marks the crossover between these two regimes. Thus, a universal scaling can be expressed in the form

$$\frac{\gamma_{Ca}}{\gamma_{Pa}} = \Psi(u) \sim \begin{cases} u^\alpha, & \text{for } \Lambda \ll \Lambda_{\text{cross}} \\ 1, & \text{for } \Lambda \gg \Lambda_{\text{cross}}, \end{cases} \quad (4)$$

where $u \equiv \Lambda/\Lambda_{\text{cross}}$. To verify the scaling above, it is important to establish a criterion for estimating $\Lambda_{\text{cross}}$. A simple way to proceed is to find a function $\Psi(u)$ that satisfies the properties mentioned above. We suggest a function of the form

$$\Psi(u) = [1 - \exp(-u)]^\alpha, \quad (5)$$

which clearly satisfies the conditions in Eq. (4) in the limits $u \gg 1$ and $u \ll 1$. Furthermore, for the three values of $\sigma_a$ shown in Fig. 2 and fixing $\Xi = 1024$, where the power law is more clearly observed, our results indicate that the effective exponent $\alpha$ is very close to unity in the small $\Lambda$ limit, as shown by the local slope analysis presented in the inset of Fig. 3 (b). A straightforward analytical proof for this regime (i.e., when $\alpha = 1$) is provided in Sec. 3 of the Supplementary Material for an emitter assumed to be planar and a spherical counter-electrode. This is consistent with the regime numerically explored here. In this way, by fixing $\alpha = 1$, it is possible to estimate an effective value of $\Lambda_{\text{cross}}$ as follows:

$$\Lambda_{\text{cross}} = \left\{ -\frac{\partial \ln[1 - \Psi(u)]}{\partial \Lambda} \right\}^{-1}. \quad (6)$$

When the derivative above is nearly constant, $\Lambda_{\text{cross}}$ is defined. Interestingly, our results show that, in this limit, $\Lambda_{\text{cross}}$ approximately corresponds to $\Xi$, for the three sharpness aspect ratios studied (see Figs. S4, S5, and S6 in Sec. 4 of the Supplementary Material). This result was confirmed when we numerically computed the value of $\Lambda_{\text{cross}}$ that best collapses the curves shown in Figs. 2 (a), (b), and (c). The results we obtained are presented in Fig. 3 (b), which represents a nearly linear behavior over at least three decades on both the vertical and horizontal axes. These findings show that the crossover between the power-law and saturation regimes is defined when $R$ is of the same order as $d_{\text{gap}}$.

Importantly, our results present a universal behavior in the sense that we managed to get a good collapse of all curves shown in Fig. 2 by normalizing the vertical variable to $\gamma_{Ca}/\gamma_{Pa}$ and the horizontal variable to $\Lambda/\Lambda_{\text{cross}}$. The collapse of the data is excellent as shown in Fig. 3 (c), which shows the results for $\sigma_a = 100$. The inset of Fig. 3 (c) confirms the collapse for all data shown in Fig. 2. This is a clear signature of scale invariance in this problem.

Therefore, our numerical results show that the ratio $\gamma_{Ca}/\gamma_{Pa}$ is solely a function of $\Lambda/\Lambda_{\text{cross}} \approx R/d_{\text{gap}}$. At this point, it is clear that when $d_{\text{gap}}$ is on the order of $R$, $\gamma_{Ca}$ is well within the crossover region and somewhat lower than the saturation value $\gamma_{Pa}$, even though $R/r_a \gg 1$. Hence, the results presented by Bonard *et al.* [16] should indicate that $\gamma_{Ca} < \gamma_{Pa}$, since $R \approx d_{gap}$. Our results shown in Fig. 3 (c) indicate that $\gamma_{Ca}/\gamma_{Pa} \approx 0.55$ when $R \approx d_{gap}$, a configuration similar to that depicted in Fig. 2 in Ref. [16], which would partially justify the apparent discrepancies observed by the authors. Of course, part of this apparent discrepancy could also be attributed to the fact that the emitter may not be a perfect HCP, thus not exactly matching the geometry simulated in our work. Nevertheless, we point out that $r_a \ll R$ is not the condition to determine that $\gamma_{Ca}$ is consistent with $\gamma_{Pa}$. Rather, $R$ must be much bigger that $d_{gap}$, which is not the case in Ref. [16].

While this work demonstrates universality for systems with curved counter-electrodes using the FEM technique, additional validation through particle-in-cell (PIC) simulations could provide further confirmation and extension of the proposed scaling laws under a broader range of conditions [33, 34]. The FEM employed in this study, however, offers distinct advantages, such as precise control over boundary conditions, enabled by advanced techniques like the MDD method [15, 30, 31], and accurate handling of geometric parameters. These features make FEM particularly well-suited for analyzing highly specific configurations, such as spherical counter-electrodes.

To conclude, our study establishes a universal scaling relationship that links the apex field enhancement factor of an HCP emitter to the curvature of the counter-electrode, expressed as

$$\gamma_{Ca} = \gamma_{Pa} \Psi(R/d_{gap}). \qquad (7)$$

This approach not only unifies observations across a range of sharpness aspect ratios ($10^2$–$10^3$), but also provides new insights into reconciling discrepancies in experimental data for systems involving curved counter-electrodes. Such universal scaling has not been explicitly demonstrated or applied in this context until now, underscoring the novelty of these findings. For example, the experimental findings by Bonard *et al.* report effective FEF values under conditions consistent with our theoretical predictions, particularly for setups involving curved counter-electrodes. Further experimental validation using similar configurations could strengthen the applicability of the proposed universal scaling relationships, especially for $r_a > 30$nm. We have recently become aware of a work by Meng *et al.* [35], which presents a remarkable analysis through an experimental approach. Their findings are based on the investigation of FE characteristics using metallic nanotips with varying emitter curvatures. While their work focuses on the curvature of the emitters, in contrast to our study, which examines the effects of counter-electrode curvature, both approaches reveal a consistent scaling behavior governed by the ratio $R/d_{gap}$. Although the sources of curvature differ, the alignment between their experimental results and our theoretical predictions highlights the universality of the proposed scaling parameter, further reinforcing its relevance across various geometrical configurations.

By establishing this universal behavior, our study contributes to a deeper understanding of the interplay between counter-electrode curvature and the apex field enhancement in FE devices. For instance, our results show that $\gamma_{Pa}$ can be extracted from an FE system with a curved counter-electrode when its radius of curvature $R$ is at least 10 times larger than $d_{gap}$. Therefore, the universality demonstrated in our results can play a role in the precise characterization of emitter parameters using a scanning anode field emission microscope (SAFEM), which typically operates with curved counter-electrodes. It can help characterize cathodes that achieve significant electron emission under relatively low applied electrostatic fields, primarily due to high local field enhancement factors arising from their geometric properties. Additionally, it enables a detailed investigation of dark current sources and supports the development of improved cathode handling techniques.

This work was supported by the Conselho Nacional de Desenvolvimento Científico e Tecnológico (CNPq), Grant No. 305688/2023-5 (TAdA).

### CONFLICT OF INTEREST

The authors have no conflict to disclose.

### DATA AVAILABILITY

The data that support the findings of this study are available from the corresponding authors upon reasonable request.

Supplementary Materials for

# Universal scaling of electrostatic effects of a curved counter-electrode on the emitter field enhancement

Thiago A. de Assis[1,2,‡] and Fernando F. Dall'Agnol [3,⋆]

[1]*Instituto de Física, Universidade Federal da Bahia,*

*Campus Universitário da Federação, Rua Barão de Jeremoabo s/n, 40170-115, Salvador, BA, Brazil*

[2]*Instituto de Física, Universidade Federal Fluminense,*

*Avenida Litorânea s/n, 24210-340, Niterói, RJ, Brazil*

[3]*Department of Exact Sciences and Education, Universidade Federal de Santa Catarina,*

*Campus Blumenau, Rua João Pessoa, 2514, Velha, Blumenau 89036-004, SC, Brazil*

[‡]thiagoaa@ufba.br

[⋆]fernando.dallagnol@ufsc.br

## 1 Orthodoxy test applied to FE measurements from single CNTs in the Bonard *et al.* work

To evaluate whether the experimental current-voltage characteristics reported by Bonard *et al.* pass the orthodoxy test, we extracted the scaled barrier parameter ($f_\text{t}$) from the experimental Fowler-Nordheim (FN) plots related to their Fig. 2 (see Ref [16] of the main text). In addition to the conditions established in the main text of this work, a FE system is considered orthodox if the current-voltage characteristics satisfy the following conditions: (i) the total system geometry remains unchanged under biasing conditions, (ii) the emission process occurs through tunneling via a Schottky-Nordheim (SN) barrier, and (iii) there is no significant voltage dependence on the emission area or the local work function. Otherwise, the field enhancement factor derived from orthodox or simplified analyses might be misleading.

The scaled barrier parameter extracted from an experimental FN plot is given by (see Ref. [19] of the main text):

$$f_\text{ext} = -\frac{s_\text{t}\eta}{S_\text{exp}(\Phi^{-1})},$$

where $s_\text{t} \approx 0.95$, which is considered a good approximation for technological purposes. The parameter $\eta = bc^2\phi^{-1/2}$, where $\phi$ represents the local work function, and $b$ and $c$ are the universal second FN constant and Schottky constant, respectively (see Table I in Ref. [19] of the main text). For $\phi = 5\,\text{eV}$, $\eta \approx 4.3989$. The term $\Phi$ typically corresponds to the midpoint value read from the horizontal axis of a nearly linear FN plot. Our orthodoxy test, based on Fig. 2 of Bonard *et al.*, yielded $f_\text{ext} = 0.22$, which is within the expected range, indicating that the FE measurements are consistent with orthodox emission behavior. We also determined the scaled barrier parameter by considering the entire data range, which lies between $f_\text{lb}$ (lower bound) and $f_\text{ub}$ (upper bound). Our results showed that $f_\text{lb} \approx 0.18$ and $f_\text{ub} \approx 0.29$, further supporting the orthodox emission across the entire emission range reported by Bonard *et al.*. Interestingly, the effective FEFs extracted from the FN plots reported by Bonard *et al.* agree, within error bars, with the results of our computational simulations conducted using the same experimental dimensions, as discussed next. Indeed, their Fig. 2 is a SEM micrograph showing a CNT with a total height of $h = 1.4\mu$m and an apex radius of curvature $r_\mathrm{a} = 7.5$ nm ($\sigma_a \approx 187$), with the rounded counter-electrode having a curvature radius of $R \approx 1\mu$m, positioned at a distance $d_\mathrm{gap} = 1.25\mu$m. Therefore, $\Xi \approx 167$ and $\Lambda \approx 133$. Analyzing the current-voltage characteristics for this system shown in their Fig. 2, Bonard *et al.* reported an effective experimental $\gamma_\mathrm{Ca} = (90 \pm 15)$. Using the same characteristic dimensions for the system, our theoretical result shows $\gamma_\mathrm{Ca} \approx 74$, which, surprisingly (due to the fact that carbon nanotubes with small radii have been explored), reveals excellent agreement within the experimental error bar.

## 2 Potential barrier for an HCP emitter

The system under investigation comprises an HCP emitter and a spherical anode. When the emitter's apex has a sufficiently small radius of curvature, the electrostatic field near its surface becomes highly non-uniform within a few nanometers of the apex. This field non-uniformity significantly influences the tunneling probability compared to scenarios with a uniform field in the same region. To explore this effect, numerical simulations were conducted to analyze the variation in the electrostatic potential along the axis normal to the emitter's apex. These simulations provide critical insights into how field non-uniformity affects electron emission. The procedure for determining the potential distribution along the axis of symmetry was systematically applied to various configurations. These configurations included emitters with apex radii of 7.5nm and 50nm, as well as sharpness aspect ratios of 100, 200, and 500. The results are shown in Figs. S1, S2 and S3.

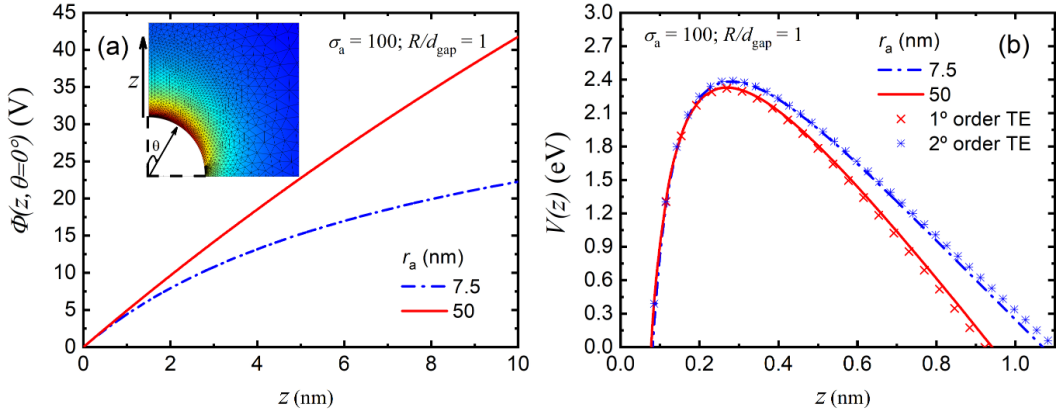

**Figure S1:** (a) The electrostatic potential numerically calculated along the azimuthal axis for an HCP emitter with $\sigma_a = 100$, $r_a = 7.5$ nm, and $r_a = 50$ nm. The coordinate $z$ is referenced from the apex position of the emitter. The inset highlights the emitter's cap, showing the mesh and the color map that reflects the intensity of the local electrostatic field. (b) The potential barrier numerically calculated for an HCP emitter with $\sigma_a = 100$, $r_a = 7.5$ nm, and $r_a = 50$ nm. An apex electrostatic field of $E_a = 5\,\text{V nm}^{-1}$ and a local work function of $5\,\text{eV}$ were used. The corresponding potential barrier, considering the first- and second-order terms in the Taylor expansion (TE) of the electrostatic potential (see Eq. (2.6) in Ref. [24] of the main text), is also presented for comparison. The results in (a) and (b) are shown for $R/d_\text{gap} = 1$, with similar trends observed for a planar counter-electrode where $R/d_\text{gap} = \infty$.

S3

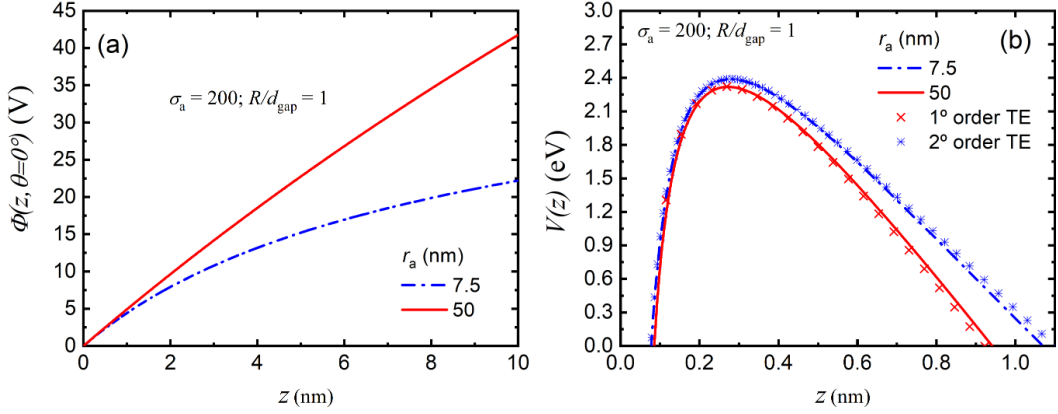

**Figure S2:** (a) The electrostatic potential numerically calculated along the azimuthal axis for an HCP emitter with $\sigma_\mathrm{a} = 200$, $r_\mathrm{a} = 7.5$ nm, and $r_\mathrm{a} = 50$ nm. The coordinate $z$ is referenced from the apex position of the emitter. (b) The potential barrier numerically calculated for an HCP emitter with $\sigma_\mathrm{a} = 200$, $r_\mathrm{a} = 7.5$ nm, and $r_\mathrm{a} = 50$ nm. An apex electrostatic field of $E_\mathrm{a} = 5$ V nm$^{-1}$ and a local work function of $5$ eV were used. The corresponding potential barrier, considering the first- and second-order terms in the Taylor expansion (TE) of the electrostatic potential (see Eq. (2.6) in Ref. [24] of the main text), is also presented for comparison. The results in (a) and (b) are shown for $R/d_\mathrm{gap} = 1$, with similar trends observed for a planar counter-electrode where $R/d_\mathrm{gap} = \infty$.

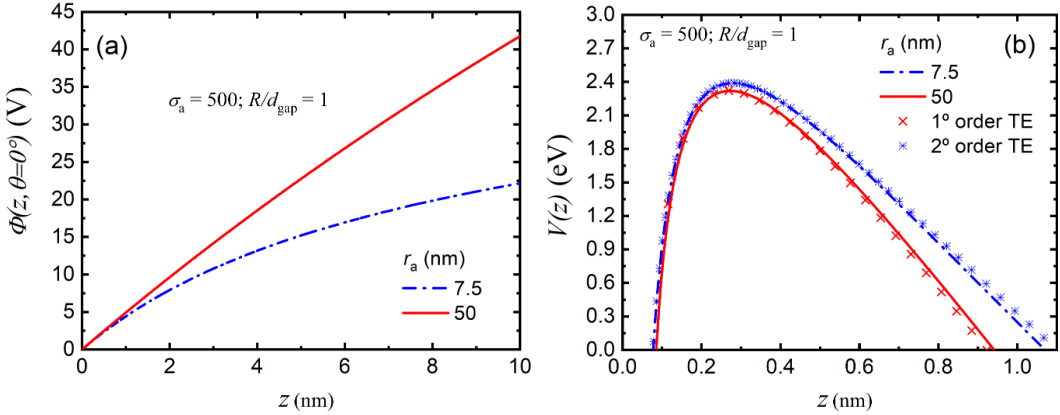

**Figure S3:** (a) The electrostatic potential numerically calculated along the azimuthal axis for an HCP emitter with $\sigma_\mathrm{a} = 500$, $r_\mathrm{a} = 7.5$ nm, and $r_\mathrm{a} = 50$ nm. The coordinate $z$ is referenced from the apex position of the emitter. (b) The potential barrier numerically calculated for an HCP emitter with $\sigma_\mathrm{a} = 500$, $r_\mathrm{a} = 7.5$ nm, and $r_\mathrm{a} = 50$ nm. An apex electrostatic field of $E_\mathrm{a} = 5$ V nm$^{-1}$ and a local work function of $5$ eV were used. The corresponding potential barrier, considering the first- and second-order terms in the Taylor expansion (TE) of the electrostatic potential (see Eq. (2.6) in Ref. [24] of the main text), is also presented for comparison. The results in (a) and (b) are shown for $R/d_\mathrm{gap} = 1$, with similar trends observed for a planar counter-electrode where $R/d_\mathrm{gap} = \infty$.

# 3 Ratio $\gamma_{Ca}/\gamma_{Pa}$ in the limit $u = R/d_{\text{gap}} \ll 1$ for a planar emitter and a spherical counterelectrode

Consider a conducting sphere with radius $R$, positioned at a distance $d_{\text{gap}}$ from a planar conductor, which is assumed to represent an emitter surface with a large apex radius. Here, $d_{\text{gap}}$ is the distance between the sphere's bottommost point and the planar emitter. The sphere is centered at $z = R + d_{\text{gap}}$, and the planar emitter is grounded ($\Phi = 0$).

Using the method of images, the sphere is represented by a point charge $q$ at $z_0 = R + d_{\text{gap}}$, and its image $-q$ at $z = -z_0$. The resulting potential in the region above the plane ($z > 0$) is given by:

$$\Phi(x, y, z) = \frac{1}{4\pi\epsilon_0} \left[ \frac{q}{\sqrt{x^2 + y^2 + (z - z_0)^2}} - \frac{q}{\sqrt{x^2 + y^2 + (z + z_0)^2}} \right]. \tag{S1}$$

Here, $\epsilon_0$ is the vacuum electric permittivity. At the planar emitter ($z = 0$), the vertical component of the electrostatic field is:

$$E_z(x, y, 0) = -\left.\frac{\partial \Phi}{\partial z}\right|_{z=0} = -\frac{q}{4\pi\epsilon_0} \left[ \frac{(z - z_0)}{(x^2 + y^2 + (z - z_0)^2)^{3/2}} - \frac{(z + z_0)}{(x^2 + y^2 + (z + z_0)^2)^{3/2}} \right]\bigg|_{z=0}. \tag{S2}$$

At the point directly below the sphere's center ($x = y = 0$):

$$E_z(0, 0, 0) = \frac{q}{2\pi\epsilon_0 z_0^2}. \tag{S3}$$

The sphere is held at a constant potential $\Phi_0$, and the potential on its surface is:

$$\Phi_{\text{sphere}} = \frac{q}{4\pi\epsilon_0 R} = \Phi_0. \tag{S4}$$

Solving for $q$, we find:

$$q = 4\pi\epsilon_0 R \Phi_0. \tag{S5}$$

Substituting this result into the Eq. (S3):

$$E_z(0, 0, 0) = \frac{4\pi\epsilon_0 R \Phi_0}{2\pi\epsilon_0 z_0^2} = \frac{2R\Phi_0}{z_0^2}. \tag{S6}$$

Since $z_0 = R + d_{\text{gap}}$, we expand $z_0^2$ in the limit $R \ll d_{\text{gap}}$, i.e.,

$$z_0^2 = (R + d_{\text{gap}})^2 \approx d_{\text{gap}}^2 + 2R d_{\text{gap}}. \tag{S7}$$

Substituting into the Eq. (S6):

$$E_z(0, 0, 0) \approx \frac{2R\Phi_0}{d_{\text{gap}}^2 + 2R d_{\text{gap}}}. \tag{S8}$$

Since in the limit $R \ll d_{\text{gap}}$, the term $2R d_{\text{gap}}$ in the denominator can be neglected. Hence,

$$E_z(0, 0, 0) \approx \frac{2R\Phi_0}{d_{\text{gap}}^2}. \tag{S9}$$

The applied electrostatic field, when the counter-electrode is curved is:

$$E_C = \frac{\Phi_0}{d_{\text{gap}}}. \tag{S10}$$

Thus, the field enhancement factor at the planar pointed emitter due to the curved counter-electrode is:

$$\gamma_{Ca} = \frac{E_z(0,0,0)}{E_C} = \frac{2R}{d_{\text{gap}}}. \tag{S11}$$

As the emitter is assumed to be planar, $\gamma_{Pa} = 1$ when $R \to \infty$. Thus, the ratio of the field enhancement factors scales as:

$$\frac{\gamma_{Ca}}{\gamma_{Pa}} \sim \frac{R}{d_{\text{gap}}}. \tag{S12}$$

## 4 Local Slope Analysis

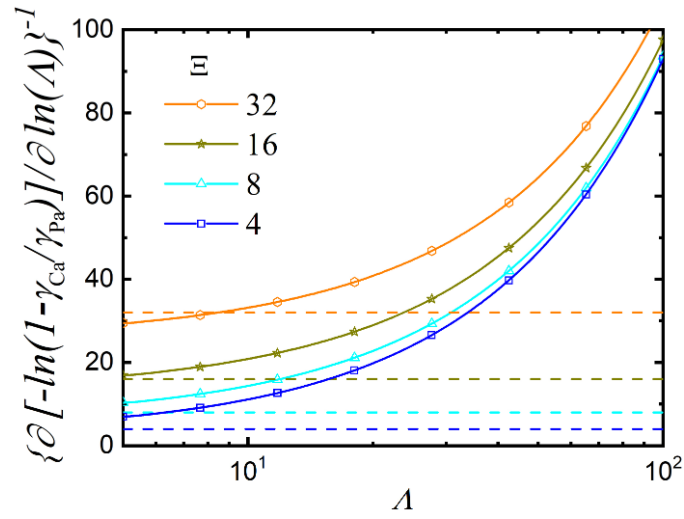

**Figure S4:** Analysis of the local slope given by Eq. (2) of the paper, using the data shown in Fig. 2(a) of the main text for $\Xi = 4, 8, 16$, and $32$ ($\sigma_\mathrm{a} = 100$). Clearly, in the limit of small values of $\Lambda$, the derivative tends to $\Xi$ (horizontal dashed lines). Similar results are observed for other sharpness aspect ratios.

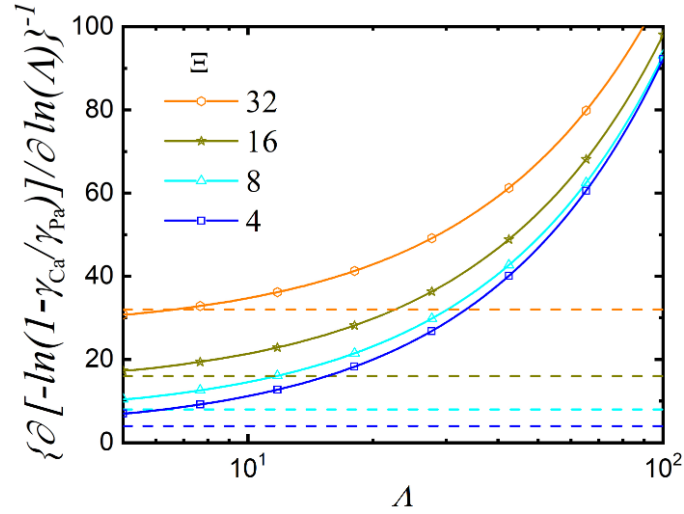

**Figure S5:** Analysis of the local slope given by Eq. (2) of the paper, using the data shown in Fig. 2(b) of the main text for $\Xi = 4, 8, 16$, and $32$ ($\sigma_\mathrm{a} = 200$). Clearly, in the limit of small values of $\Lambda$, the derivative tends to a constant close to $\Xi$ (horizontal dashed lines). Similar results are observed for other sharpness aspect ratios.

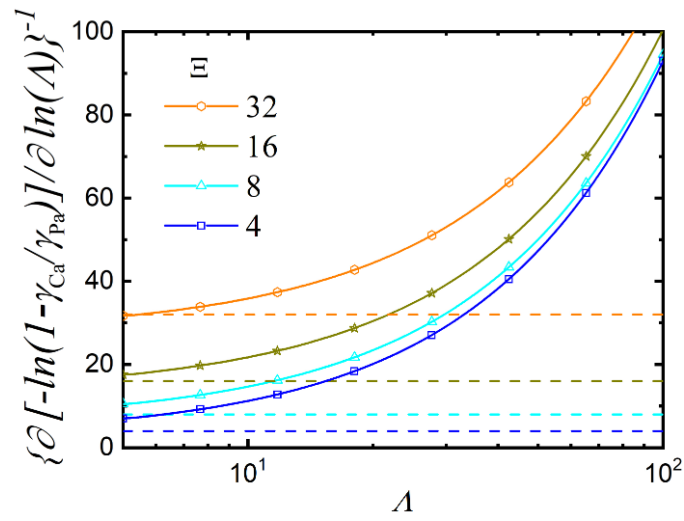

**Figure S6:** Analysis of the local slope given by Eq. (2) of the paper, using the data shown in Fig. 2(c) of the main text for $\Xi = 4, 8, 16$, and $32$ ($\sigma_a = 500$). Clearly, in the limit of small values of $\Lambda$, the derivative tends to a constant close to $\Xi$ (horizontal dashed lines). Similar results are observed for other sharpness aspect ratios.